\documentclass[twocolumn,showpacs,floatfix,aps]{revtex4}
\usepackage{graphicx}
\usepackage{bm}

\begin{document}

\title{Why only half of the fermionic atoms were converted to molecules
    by a Feshbach resonance?}
\author{Hui Hu${}^{1}$, Feng Yuan${}^{1,2}$, and Shaojin Qin${}^{1,3}$}
\affiliation{ ${}^{1}$Abdus Salam International Center for
Theoretical Physics, P. O. Box 586, Trieste 34100, Italy\\
${}^{2}$Department of Physics, Qingdao University, Qingdao 266071, China \\
${}^{3}$Institute of Theoretical Physics, Chinese Academy of
Sciences, Beijing 100080, China }
\date{\today}

\begin{abstract}
In some recent experiments an ultracold gas of $^{40}$K$_2$
(or $^{6}$Li$_2$) molecules has been produced from a degenerate
two-component Fermi gas of $^{40}$K (or $^{6}$Li) atoms by
adiabatic passage through a Feshbach resonance. The maximum
atom-molecule transfer efficiency is reported to be about
$50\%$. We propose a simple microscopic model to characterize the
ground state of the gas in the vicinity of the resonance, and show
that the term describing the atom-molecule coupling is responsible
for the observed efficiency $50\%$. Our result also suggests that
the experiments have produced a molecular condensate.
\end{abstract}

\maketitle

Recently the group at the Joint Institute for Laboratory
Astrophysics (JILA) has succeeded in converting an ultracold Fermi
gas of $^{40}$K atoms into an ultracold gas of $^{40}$K$_2$
molecules by sweeping a magnetic field across a Feshbach resonance
\cite{jin}. Using a similar method with fermionic $^6$Li atoms,
several experiments have also reported the production of
$^6$Li$_2$ molecules \cite{hulet,ens,jochim}. This impressive
experimental achievement of ultracold molecules from fermionic
atoms not only raises the prospect of creating novel molecular
Bose-Einstein condensates, but also sparks a hot race to realize
the Bardeen-Cooper-Schrieffer (BCS) transition to superfluid phase
of the fermionic components \cite {julienne,cho}, in which the
superfluid transition temperature is predicted to be maximum near
a Feshbach resonance\cite{pla,holland,griffin}.

At present, several of the experimental results are not well
understood. At first, a common feature observed in these
experiments is that the maximum atom-molecule transfer efficiency
is limited to $\sim 50\%$ at low temperatures
\cite{jin,hulet,jochim}. This observation is not within the
prediction of the well-known Landau-Zener theory
\cite{landau,zener}, and therefore constitutes a theoretical
challenge. The Landau-Zener model, which is expected to be
applicable in this case, predicts that the transfer probability is
proportional to $1-\exp (-\dot{B}^{-1}),$ when the magnetic field
is ramped from a field far above the Feshbach resonance to the one
far below, with $\dot{B}^{-1}$ being the inverse ramp rate.
Actually, the
measured fraction of fermionic atoms remaining in the gas as a function of $%
\dot{B}^{-1}$ exhibits the predicted exponential dependence.
However, the model does not predict that the conversion should be
$50\%$ at the resonance when the ramp rate is sufficiently slow.
What measured in the experiments thus can not be comprehended by
the Landau-Zener theory. Another major problem of interest is
whether the molecular gas is actually a molecular Bose-Einstein
condensate. In the JILA experiment, the number of the molecules
produced is very large \cite{jin}. If one assumes that the
molecules have a temperature comparable to that of the fermionic
atoms from which they were made, the number of molecules is in
fact already large enough to imply that a molecular Bose-Einstein
condensate might be formed \cite{jin}.

The scope of the present paper is to give the simplest possible physical
picture of an atom-molecule mixture in close proximity to the Feshbach
resonance, essentially at zero temperature. The model is not made to give
quantitative predictions, but, as we shall see, it qualitatively reproduces
the experimental observations of the maximum atom-molecule transfer
efficiency of $\sim 50\%$, and, of course, it is consistent with the
expectation that the experiments have indeed produced a molecular condensate.

{\it The model}: --- Our starting point is a spatially homogeneous gas of
the atom-molecule mixture, in which the molecules are created from a
degenerate two-component Fermi gas by adiabatic passage through a Feshbach
resonance. The model system in the grand-canonical assemble can then be
qualitatively described by \cite{pla,holland,griffin}
\begin{eqnarray}
{\cal H-}\mu {\cal N} &=&\sum\limits_{{\bf k}\sigma }\left( \epsilon _{{\bf k%
}}^0-\mu \right) c_{{\bf k}\sigma }^{+}c_{{\bf k}\sigma }+U_{bg}\sum\limits_{%
{\bf kk}^{\prime }}c_{{\bf k}\uparrow }^{+}c_{-{\bf k}\downarrow }^{+}c_{-%
{\bf k}^{\prime }\downarrow }c_{{\bf k}^{\prime }\uparrow }  \nonumber \\
&&+\sum\limits_{{\bf q}}\left( E_{{\bf q}}^0+2\nu -2\mu \right) b_{{\bf q}%
}^{+}b_{{\bf q}}  \nonumber \\
&&+g\sum\limits_{{\bf kq}}\left( b_{{\bf q}}^{+}c_{{\bf k+q}/2\uparrow }c_{-%
{\bf k+q}/2\downarrow }+h.c.\right) .  \label{fullhami}
\end{eqnarray}
Here we describe the two components of fermionic atoms as pseudospins $%
\sigma =\uparrow ,\downarrow $. The operators $c_{{\bf k}\sigma }$ and $b_{%
{\bf q}}$ represent, respectively, the annihilation operator of a
fermion atom with the kinetic energy $\epsilon _{{\bf
k}}^0=\hbar^{2}{\bf k}^2/2m$ and a
molecular $b$ boson with the energy spectrum $E_{{\bf q}}^0+2\nu =\hbar^{2}{\bf q}%
^2/4m+2\nu $. The detuning or the binding energy of the Bose
molecules is denoted by $2\nu $ and can be conveniently adjusted
by an external magnetic field. The background fermion-fermion
interaction $U_{bg}$ originates from the nonresonant processes,
while the $g$ is a coupling constant associated with the Feshbach
resonance. The chemical potential $\mu $ is used to impose
the conversion of the bare fermionic atoms: $\left\langle {\cal N}%
\right\rangle =\left\langle \sum_{{\bf k}\sigma }c_{{\bf k}\sigma }^{+}c_{%
{\bf k}\sigma }\right\rangle +2\left\langle \sum_{{\bf q}}b_{{\bf q}}^{+}b_{%
{\bf q}}\right\rangle =N.$

The Hamiltonian (\ref{fullhami}) has been suggested to be valid at
all detunings $\nu $ \cite{holland}. In this paper, however, we
are only interested in the {\em vicinity} of the Feshbach
resonance, in which $\nu \approx 0$.  The dominant term in this
region is the last term in Eq. (\ref{fullhami}) describing the
atom-molecule coupling.
We wish to show that it {\em %
alone} is responsible for the observed maximum transfer efficiency
of $\sim 50\%$. For this purpose, we therefore write down a
truncated effective Hamiltonian in the vicinity of the Feshbach
resonance
\begin{eqnarray}
{\cal H}_{eff}{\cal -}\mu {\cal N} &=&\sum\limits_{{\bf kq}}\left( b_{{\bf q}%
}^{+}c_{{\bf k+q}/2\uparrow }c_{-{\bf k+q}/2\downarrow }+h.c.\right)
\nonumber \\
&&-\mu \left( \sum\limits_{{\bf k}\sigma }c_{{\bf k}\sigma }^{+}c_{{\bf k}%
\sigma }+2\sum\limits_{{\bf q}}b_{{\bf q}}^{+}b_{{\bf q}}\right) ,
\label{effhami}
\end{eqnarray}
where the coupling constant $g=1$ hereafter. The model is still too
complicated to be exactly solvable. To proceed, we further restricted
ourself at {\em zero} temperature in accordance with the expectation of the
appearance of the molecular Bose-Einstein condensation. The $b$ bosons are
then replaced by a Bose condensate in the ${\bf q=0}$ state, described by $%
\left\langle b_{{\bf q=0}}\right\rangle =\Phi _m=\left| \Phi _m\right|
e^{i\gamma }$. Within this simple mean-field approximation, the Hamiltonian $%
{\cal H}_{eff}{\cal -}\mu {\cal N}$ can be easily diagonalized by
constructing the Bogoliubov quasiparticles according to the
general canonical transformation \cite{bogo,valatin}
\begin{equation}
\left(
\begin{array}{c}
\alpha _{{\bf k}\uparrow } \\
\alpha _{-{\bf k}\downarrow }^{+}
\end{array}
\right) =\left(
\begin{array}{cc}
\cos \theta & -e^{i\gamma }\sin \theta \\
e^{i\gamma }\sin \theta & \cos \theta
\end{array}
\right) \left(
\begin{array}{c}
c_{{\bf k}\uparrow } \\
c_{-{\bf k}\downarrow }^{+}
\end{array}
\right) ,  \label{bogoliubov}
\end{equation}
where the transformation angle is specified as $2\theta =\pi /2+\arctan (\mu
/\left| \Phi _m\right| )$. The resulting many-body Hamiltonian now becomes
diagonal in the fermion operators
\begin{eqnarray}
{\cal H}_{eff}{\cal -}\mu {\cal N} &=&-2\mu \left| \Phi _m\right|
^2+\nonumber \\
&&\sum\limits_{{\bf k}}\left[ \epsilon _{{\bf k}}\left( \alpha _{{\bf k}%
\uparrow }^{+}\alpha _{{\bf k}\uparrow }+\alpha _{{\bf
k}\downarrow }^{+}\alpha _{{\bf k}\downarrow }-1\right) -\mu
\right]  \label{effhami2}
\end{eqnarray}
with $\epsilon _{{\bf k}}=(\mu ^2+\left| \Phi _m\right| ^2)^{1/2}$ being the
corresponding quasiparticle spectrum. Hence at zero temperature the ground
state energy is obtained as
\begin{equation}
E_{gs}=-2\mu \left| \Phi _m\right| ^2-M\left( \mu +(\mu ^2+\left| \Phi
_m\right| ^2)^{1/2}\right) ,  \label{egs}
\end{equation}
where $M=\sum_{{\bf k}}1$ represents the number of available states for the
fermionic atoms. The saddle point equations are derived by minimizing the
ground state energy with respect to $\left| \Phi _m\right| ^2,$
\begin{equation}
\frac{\partial E_{gs}}{\partial \left| \Phi _m\right| ^2}=-2\mu -\frac M{%
2(\mu ^2+\left| \Phi _m\right| ^2)^{1/2}}=0,  \label{s1}
\end{equation}
and by requiring the conversion of the number of atoms, $-\partial
E_{gs}/\partial \mu =N$, or
\begin{equation}
2\left| \Phi _m\right| ^2+M+\frac{\mu M}{(\mu ^2+\left| \Phi
_m\right| ^2)^{1/2}}=N.  \label{s2}
\end{equation}
By solving those equations, we can determine the occupation of the molecules
and the chemical potential as a function of $N/M$.

{\it Discussion}: --- We first consider the limit $N/M\ll 1$ that
may correspond to the real experimental situation of dilute atomic
gases. In this limit, Taylor-expanding the saddle point equations
in powers of $\left| \Phi _m\right| ^2/M$ and $N/M$ yields to the
solution (the factor $g$ has been restored)
\begin{eqnarray}
2\left| \Phi _m\right| ^2 &=&N/2,  \label{mr1} \\
\mu &=&-\frac{gM^{1/2}}2\left( 1-\frac{\left| \Phi _m\right|
^2}M\right) , \label{mr2}
\end{eqnarray}
in leading order. This is the main result in this work. The
obtained fraction of the molecules in the mixture, $2\left| \Phi
_m\right| ^2/N=0.5$, is in excellent agreement with the
experimental result by Jochim and co-workers, that is, either a
pure atomic or molecular sample relaxes to an atom-molecule
mixture with equal fraction \cite{jochim}. On the other hand, it
should also be compared with the experimental observations of the
maximum atom-molecule transfer efficiency of $\sim 50\%$
\cite{jin,hulet,jochim}. As seen in the experiments, the molecules
are predominantly formed when the magnetic field traverses the
Feshbach resonance peak \cite {jin,hulet,ens,jochim}. Due to the
sufficiently slow ramp of the magnetic field, the fraction of the
molecules in the ground state of the mixture is in fact equivalent
to the maximum atom-molecule transfer efficiency, if we don't care
about the small loss due to inelastic processes.

\begin{figure}[tbp]
\centerline{\includegraphics[width=5.5cm,angle=-90,clip=]{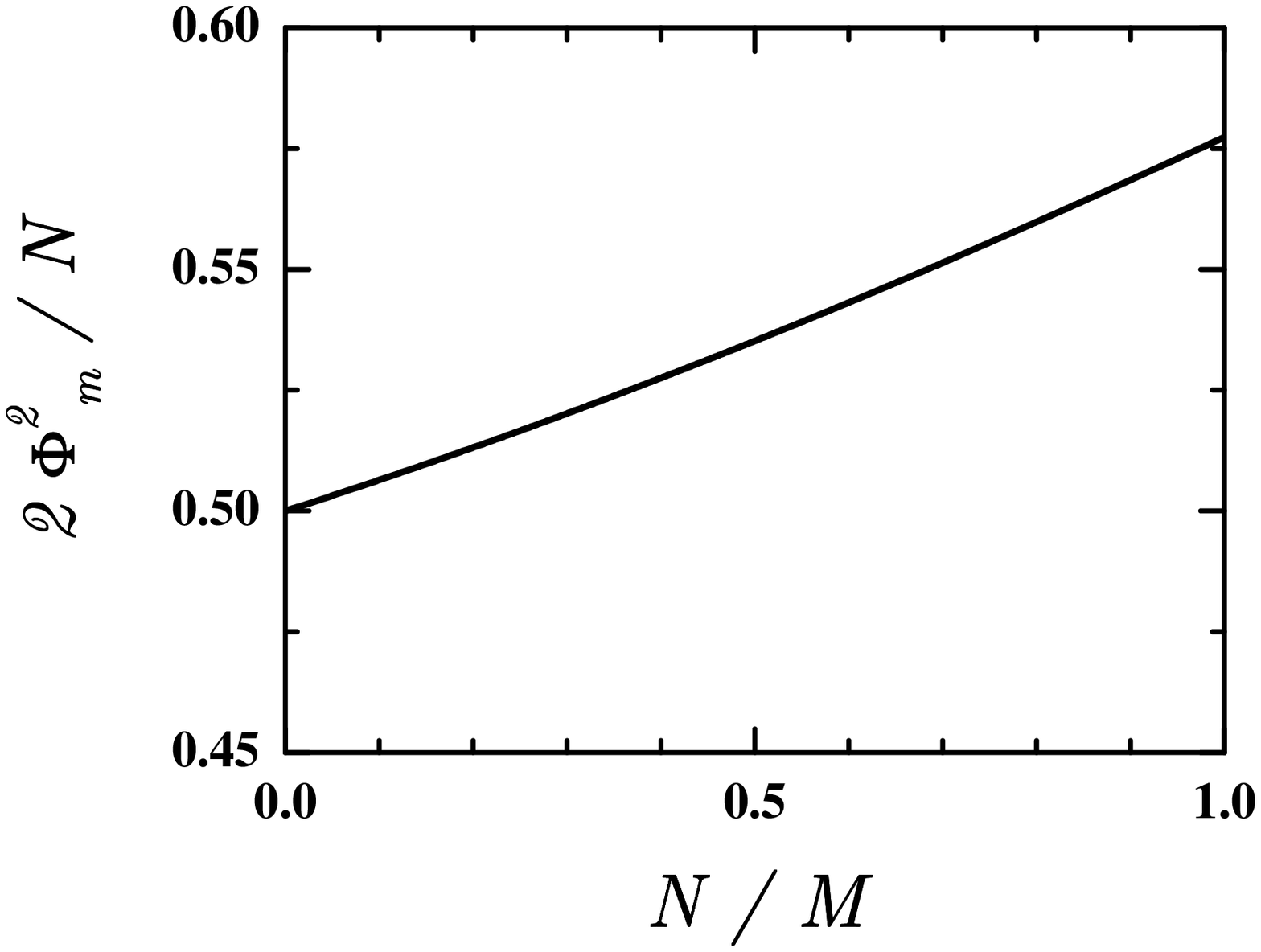}}
\caption{The fraction of the molecules of a homogeneous
atom-molecule mixture, as a function of $N/M$. In the dilute limit
of $N/M\ll 1$, the calculated fraction of the molecules is in
excellent agreement with the maximum atom-molecule transfer efficiency of $%
\sim 50\%$, as observed in some recent experiments
\cite{jin,hulet,jochim}.} \label{fig1}
\end{figure}

\begin{figure}[tbp]
\centerline{\includegraphics[width=5.5cm,angle=-90,clip=]{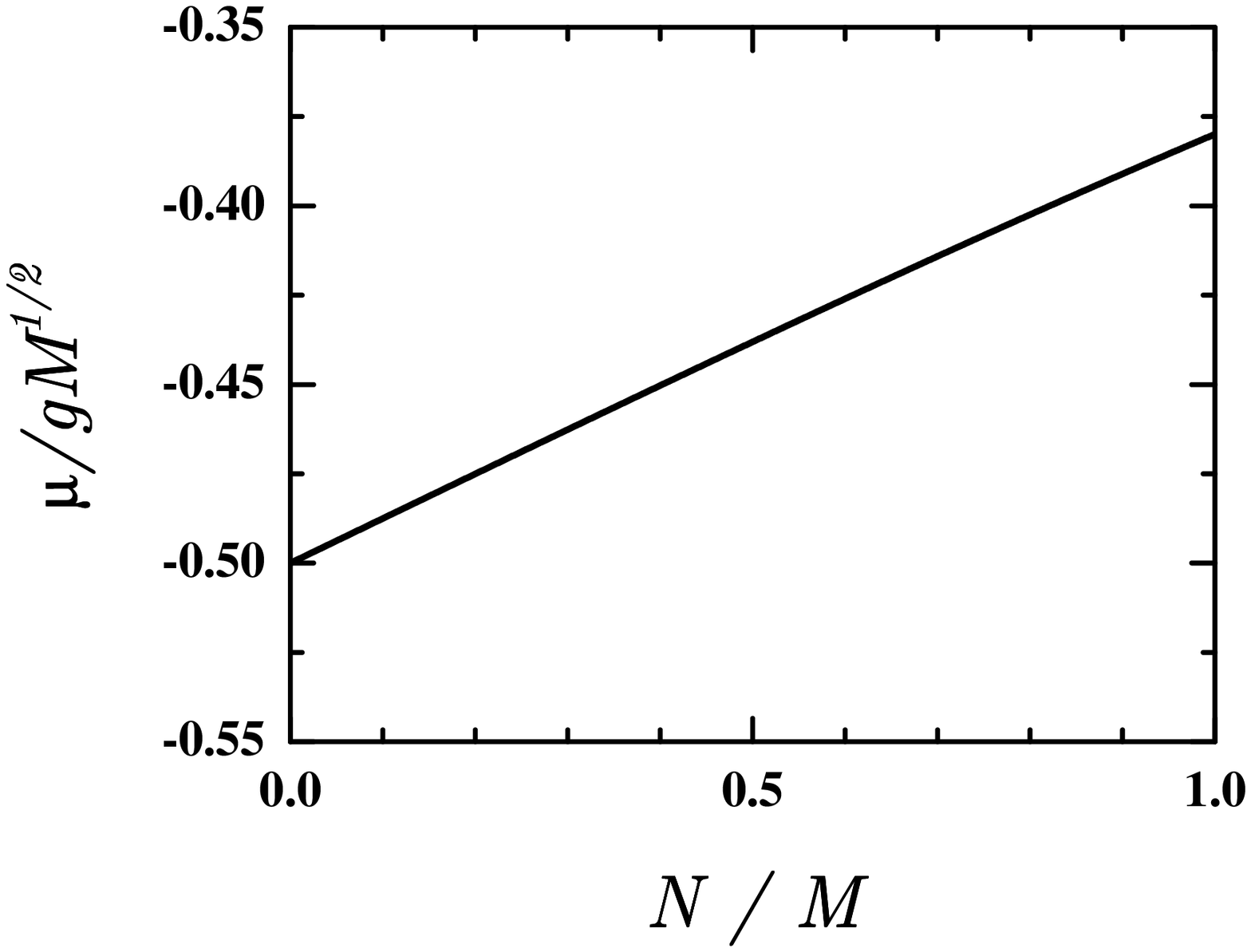}}
\caption{The chemical potential of a homogeneous atom-molecule
mixture, as a function of $N/M$.} \label{fig2}
\end{figure}

Another important feature of our model is that the chemical
potential is pinned at
\begin{equation}
\mu =-gM^{1/2}/2=-\tilde{g}/2  \label{miu}
\end{equation}
in the dilute limit of $N/M\ll 1$, where $\tilde{g}=gM^{1/2}$ is
independent on the volume $V$, since $g\propto V^{-1/2}$ and $%
M\propto V$. This pinning can be qualitatively understood by
considering a two-particle mixture, with either two atoms or one
molecule, described by an
effective Hamiltonian
\begin{equation}
{\cal H}_{eff}=\tilde{g}\sum_{{\bf i}}\left( b_{{\bf %
i}}^{+}c_{{\bf i}\uparrow }c_{{\bf i}\downarrow }+h.c.\right)
\label{effhami3}
\end{equation}
in real space. The mixture has two states: $\left|
2_A,0_M\right\rangle $ or $\left| 0_A,1_M\right\rangle $. On the
basis of these two states, the Hamiltonian takes the form,
\begin{equation}
{\cal H}_{eff}=\left(
\begin{array}{ll}
0 & \tilde{g} \\
\tilde{g} & 0
\end{array}
\right) .  \label{twolevel}
\end{equation}
Its ground state, which is the bond state of $\left|
2_A,0_M\right\rangle $ and $\left| 0_A,1_M\right\rangle $, can be
nicely regarded as a composite molecule with energy $-\tilde{g}$.
In the dilute limit, the ground state of the $N$-particle mixture
may therefore be viewed as a non-interacting gas with $N/2$ such
kind of composite molecules, distributed uniformly in the real
space. This scenario is agreement with the half populated fraction
of the molecules. Adding or removing a particle from the mixture
thus costs energy $\tilde{g}/2$, which is the precise value of the
chemical potential in Eq. (\ref{miu}).

The full solution of the saddle point equations as a function of
$N/M$ is shown in Figs. (1) and (2). The faction of the molecules
is only weakly dependent on the value of $N/M$. We believe that,
it will also be robust against the inclusion of the kinetic and
fermion-fermion interaction terms, {\it{i.e.}}, the terms in the
first and second lines of the full Hamiltonian (\ref{fullhami}),
which we have treated as irrelevant perturbations in the vicinity
of the resonance \cite{note}.

Further, analogous to the Bose-Einstein condensates with spin
\cite{hotl,wu}, more profound structures may be expected for
the atom-molecule mixture.
Besides the condensation of the molecules, the fermionic atoms
may also acquire an off-diagonal long range order since the
resonance induced an attractive pair interaction in form of
$c_{i\downarrow}^{+}c_{i\uparrow}^{+} c_{j\uparrow}c_{j\downarrow}$
among the atoms.  The interplay between the phase of the molecule
condensate and the phase of the off-diagonal long range order
for atoms is a new problem for both experimental and theoretical
studies.

The results presented so far is only valid at zero temperature, at
which the mixture is strongly degenerate. The fraction of the
molecules, of course, will be qualitatively affected by the
inclusion of a finite temperature. This is the case in the ENS
experiment \cite{ens}, in which the atomic component in the gas is
no longer quantum degenerate \cite{kss} and therefore
atom-molecule transfer efficiency can reach $85\%$. A theoretical
investigation of the effective Hamiltonian (\ref{effhami}) at
finite temperature is highly non-trivial, since it requires the
treatment beyond the mean-field approximation.
We will discuss these points elsewhere.

In conclusion, we have proposed a simple physical model to
characterize the essential physics of an atom-molecule mixture in
close proximity to the Feshbach resonance at zero temperature. The
fraction of the molecules obtained from this model agrees well
with the maximum atom-molecule transfer efficiency of $\sim 50\%$,
as observed in some recent experiments \cite{jin,hulet,jochim}.
This excellent agreement also justifies our mean-field treatment
for the molecule operator, and suggests that the experiments have
indeed produced a molecular condensate.


\begin{thebibliography}{99}
\bibitem{note} Within the mean-field approximation, these perturbations can
be easily treated. However, to do so, one has to introduce an
additional energy scale $\epsilon _L=\hbar ^2/2mL^2$ with $L$
being the characteristic size of the system, which can hardly be
determind from the experiments.
\end{thebibliography}
\end{document}